\documentclass[12pt]{article}
\usepackage{graphicx}
\begin{document}
\centerline{\bf Simulation of geographical trends in Chowdhury ecosystem model}

\bigskip
Klaus Rohde* and Dietrich Stauffer**

\bigskip
\noindent
*Zoology, University of New England, Armidale NSW 2351, Australia.

\bigskip
\noindent
**Instituto de F\'{\i}sica, Universidade
Federal Fluminense; Av. Litor\^{a}nea s/n, Boa Viagem,
Niter\'{o}i 24210-340, RJ, Brazil; visiting from 
Institute for Theoretical Physics, Cologne University, D-50923 K\"oln,
Euroland.

\bigskip
e-mail: krohde@metz.une.edu.au, stauffer@thp.uni-koeln.de
\bigskip

Abstract: A computer simulation based on individual births and deaths gives
a biodiversity increasing from cold to warm climates, in agreement with 
reality. Complexity of foodwebs increases with time and at a higher rate at
low latitudes, and there is a higher rate of species creation at low
latitudes. Keeping many niches empty makes the results correspond
more closely to natural gradients.

\bigskip
\section{Introduction}

Chowdhury et al \cite{csk} have proposed a model, recently reviewed in 
\cite{Chowdhury,Stauffer}, that aims to 
model microevolution (at the level of populations over ecological 
time) and macroevolution (at the level of species and above, over 
geological time). 
So far, geographical trends have not been included in the model. However, 
evolution is not uniform across all habitats and regions, but shows 
distinct geographical trends. Most distinct and best studied are 
latitudinal gradients in species diversity, i.e., an increase in 
species numbers from high to low latitudes, e.g. from the Arctic to the 
Amazonas. A realistic model must 
take such gradients into account, and the possibility of 
incorporating gradients can indeed be used as a test for the 
usefulness of the model.

Whereas evidence for latitudinal gradients is convincing, empirical 
evidence for greater productivity at low latitudes is equivocal. We 
examine this question using animal numbers as rough indicators of 
productivity.

Furthermore, fossil data indicate that diversity has increased over 
geological time, e.g., \cite{Jablonski99}, and ecosystems of different 
age differ distinctly in diversity. Thus, the Indo-Pacific Ocean is 
much older (as old as oceans themselves, i.e. many hundred million 
years) than the Atlantic which began to form only about 150 million 
years ago. Diversity of teleost fish and many other groups is much 
greater in the former. Also, although latitudinal gradients are found 
in both oceans, they are much more distinct in the Indo-Pacific 
\cite{Rohde93}. A further test for the validity of a model would 
therefore be whether such differences can be quantitatively reproduced by it.

Rohde \cite{Rohde92,Rohde05} has suggested that latitudinal gradients in 
diversity arise as the result of faster evolutionary rates at high 
temperatures. One implication of this hypothesis is that new species, 
many with great complexity, arise predominantly in the tropics, from 
where they spread into higher latitudes (also \cite{Jablonski93}). 
Moreover, taxa with little dispersal/vagility should have steeper 
gradients than those with great dispersal/vagility.

We adapt the parameters in the model to test the following:  1) 
species numbers decrease from low to high latitudes at rates 
corresponding to empirical data; 2)  complexity of ecosystems, as indicated by 
the number of trophic
levels, increases over evolutionary time and at a higher rate in the
tropics; 3) new species arise predominantly 
at low latitudes, 4) there is an increase of productivity (indicated 
by total animal numbers) from high to low latitudes; 5) diversity 
gradients for taxa that diffuse slowly are steeper than those for 
taxa that diffuse more rapidly into higher latitudes.

\begin{figure}[hbt]
\begin{center}
\includegraphics[angle=-90,scale=0.5]{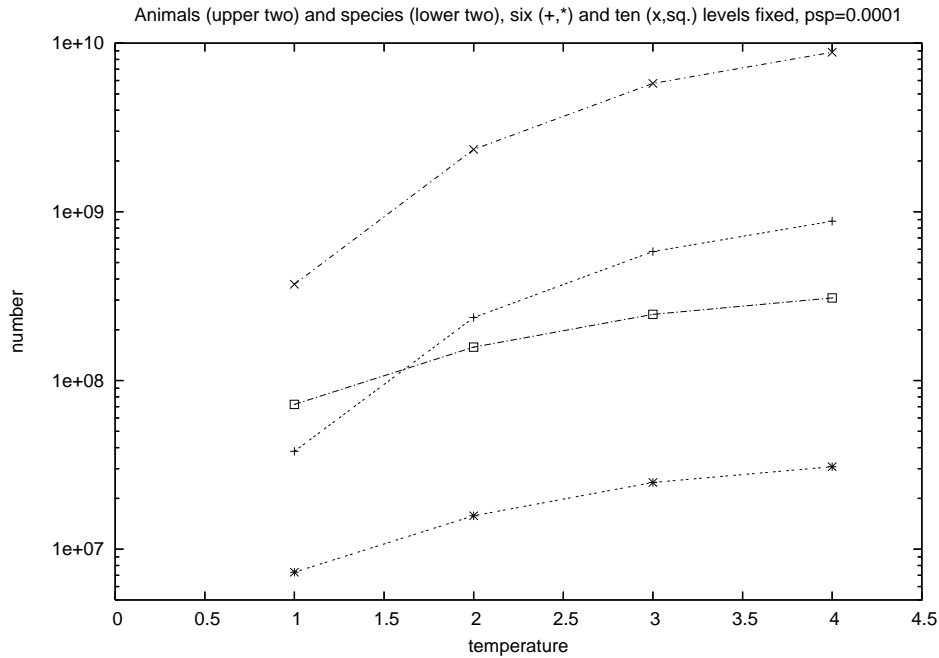}
\end{center}
\caption{
Summed numbers of animals (+ and x) and species (stars and squares) in 20 
million iterations, versus temperature (= line number in $4 \times 4$ lattice). 
}
\end{figure}

\begin{figure}[hbt]
\begin{center}
\includegraphics[angle=-90,scale=0.32]{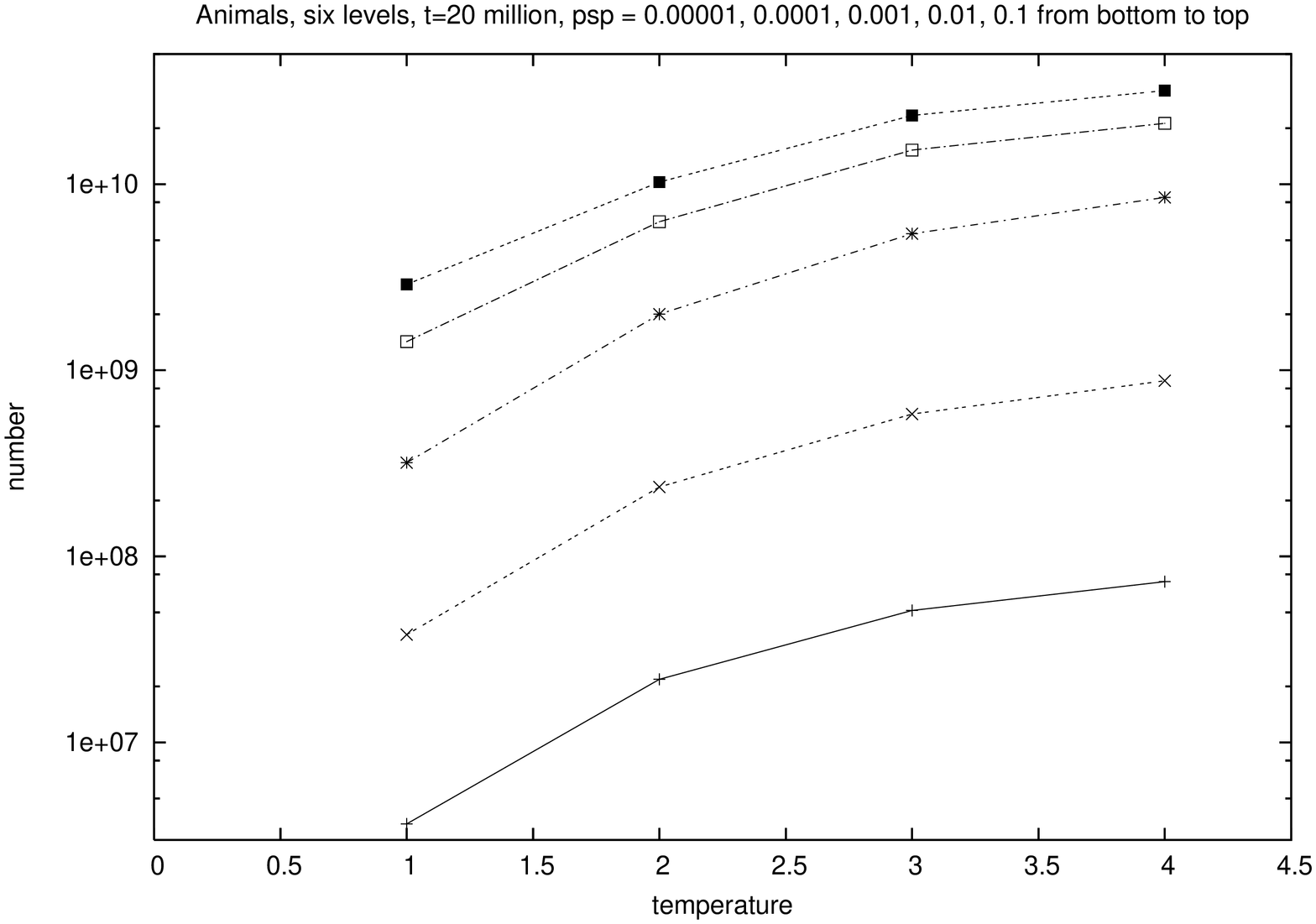}
\includegraphics[angle=-90,scale=0.32]{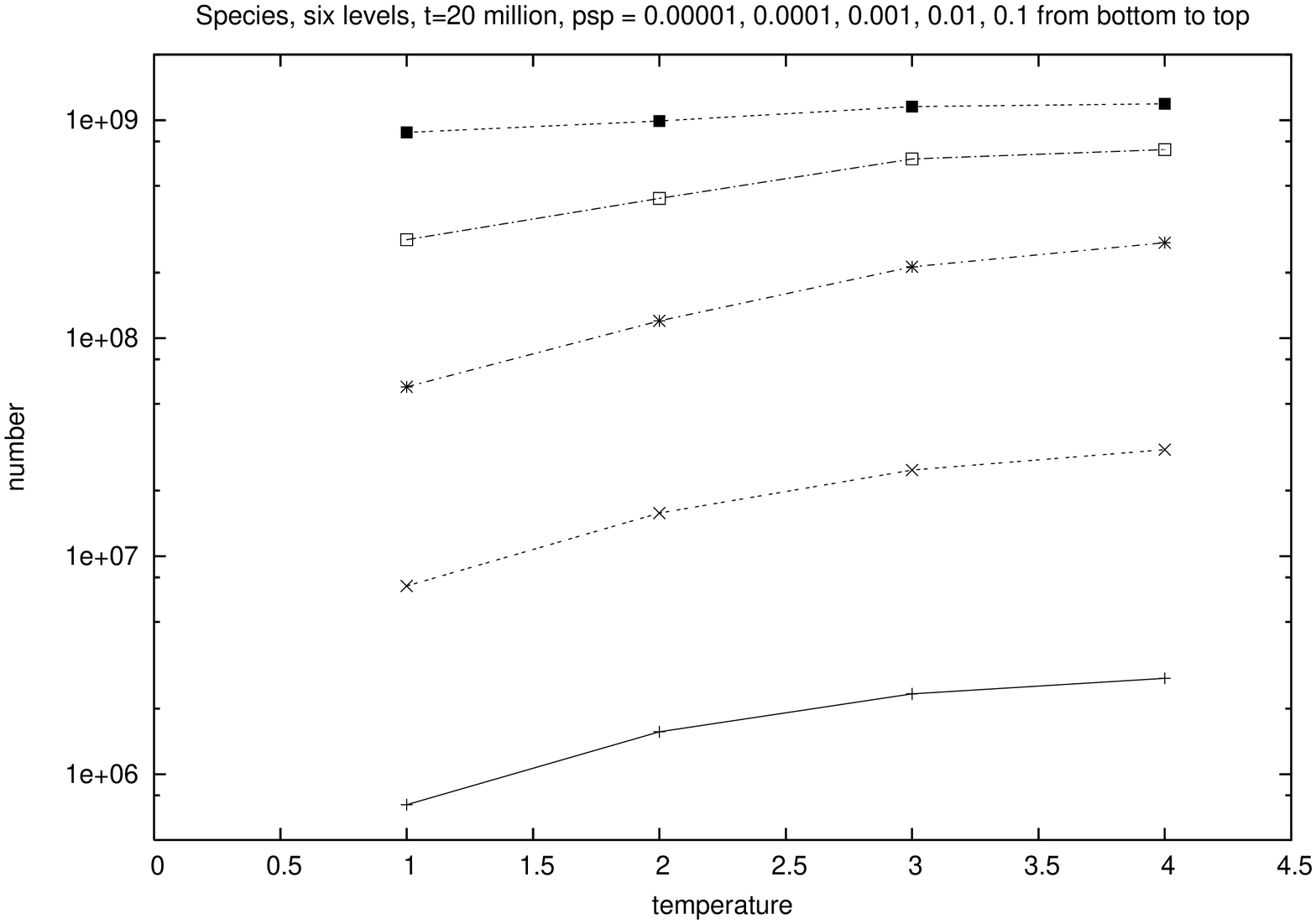}
\end{center}
\caption{
Summed number of animals (part a) and species (part b) for speciation 
probabiltiy $10^{-5} \dots 10^{-1}$ increasing from bottom to top. 
}
\end{figure}

\begin{figure}[hbt]
\begin{center}
\includegraphics[angle=-90,scale=0.5]{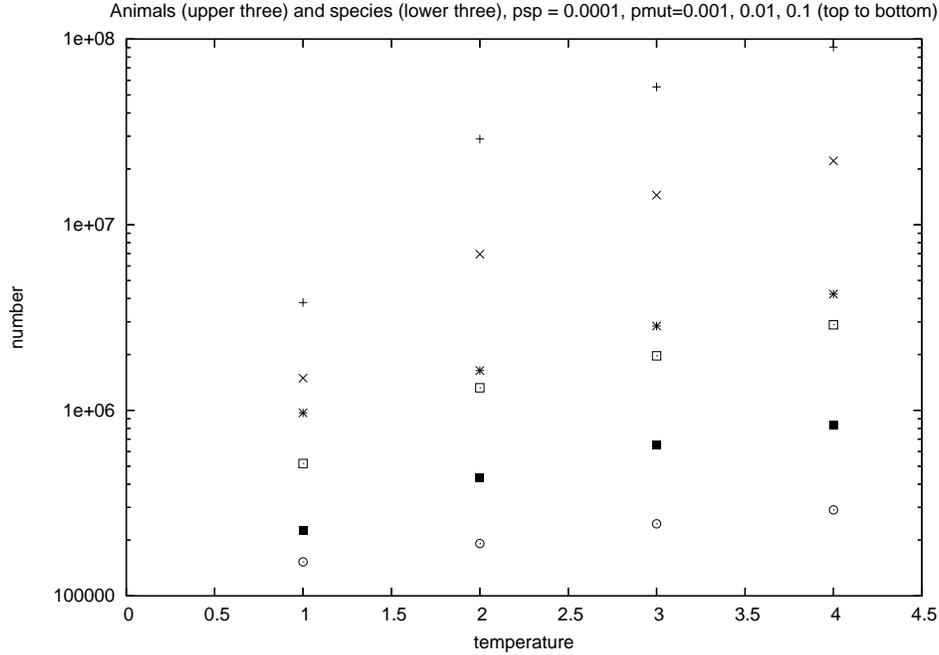}
\end{center}
\caption{
Summed animal numbers (upper data) and species numbers (lower data) for
mutation probability 0.1, 0.01, and 0.001 (from top to bottom); $p_{sp} =
0.0001$.
}
\end{figure}

\begin{figure}[hbt]
\begin{center}
\includegraphics[angle=-90,scale=0.32]{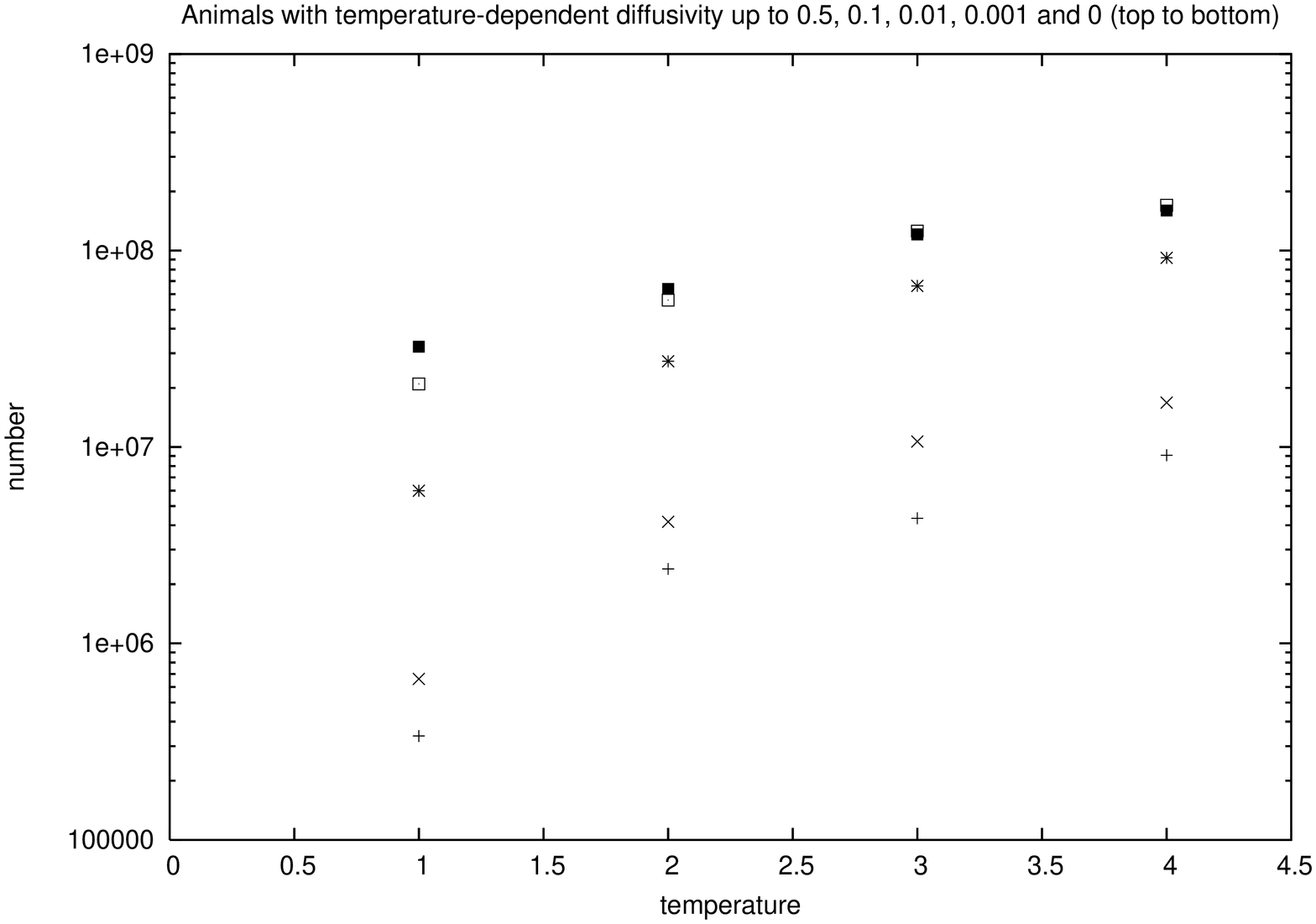}
\includegraphics[angle=-90,scale=0.32]{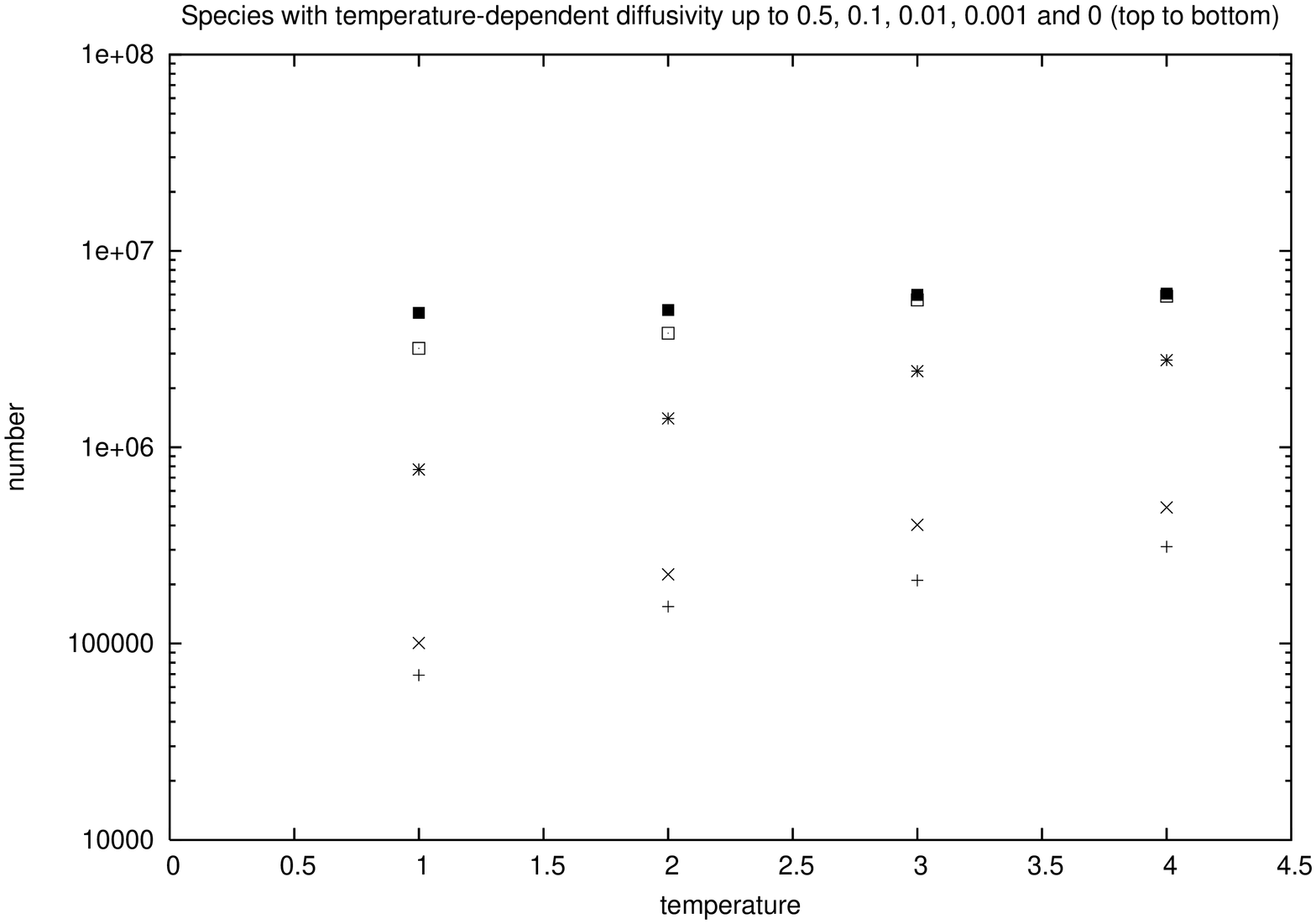}
\end{center}
\caption{
Effect of diffusion for animals (part a) and species (part b). 
 }
\end{figure}

\begin{figure}[hbt]
\begin{center}
\includegraphics[angle=-90,scale=0.5]{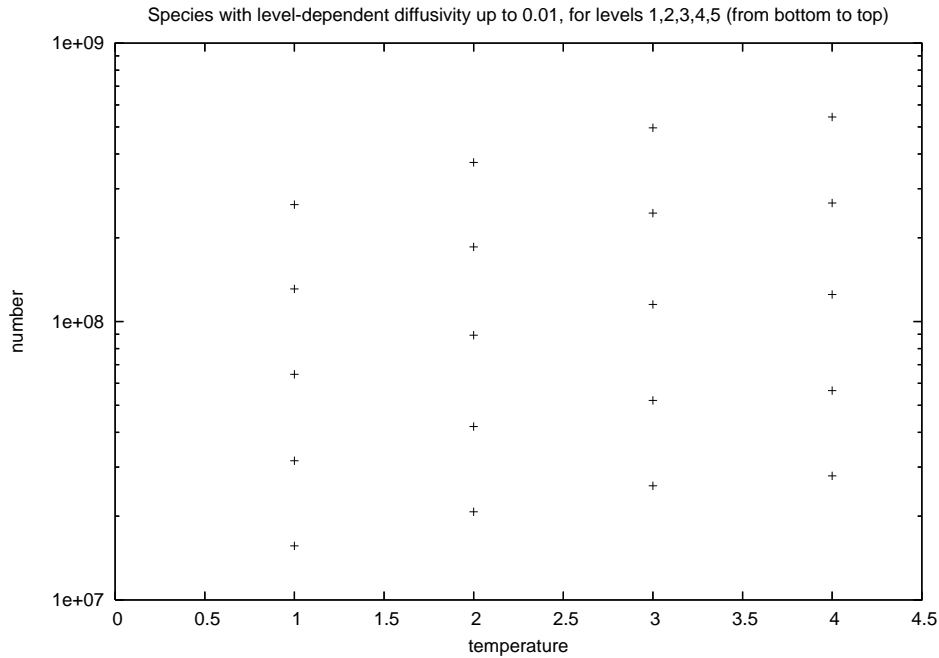}
\end{center}
\caption{
Summed species numbers, for five of the six simulated levels, when the
diffusivity is 0.01 for the top food level and decreases linearly with 
increasing level number to zero at level six. The low levels correspond to
the large numbers; the top level contains at most one species at a time.
}
\end{figure}

\begin{figure}[hbt]
\begin{center}
\includegraphics[angle=-90,scale=0.31]{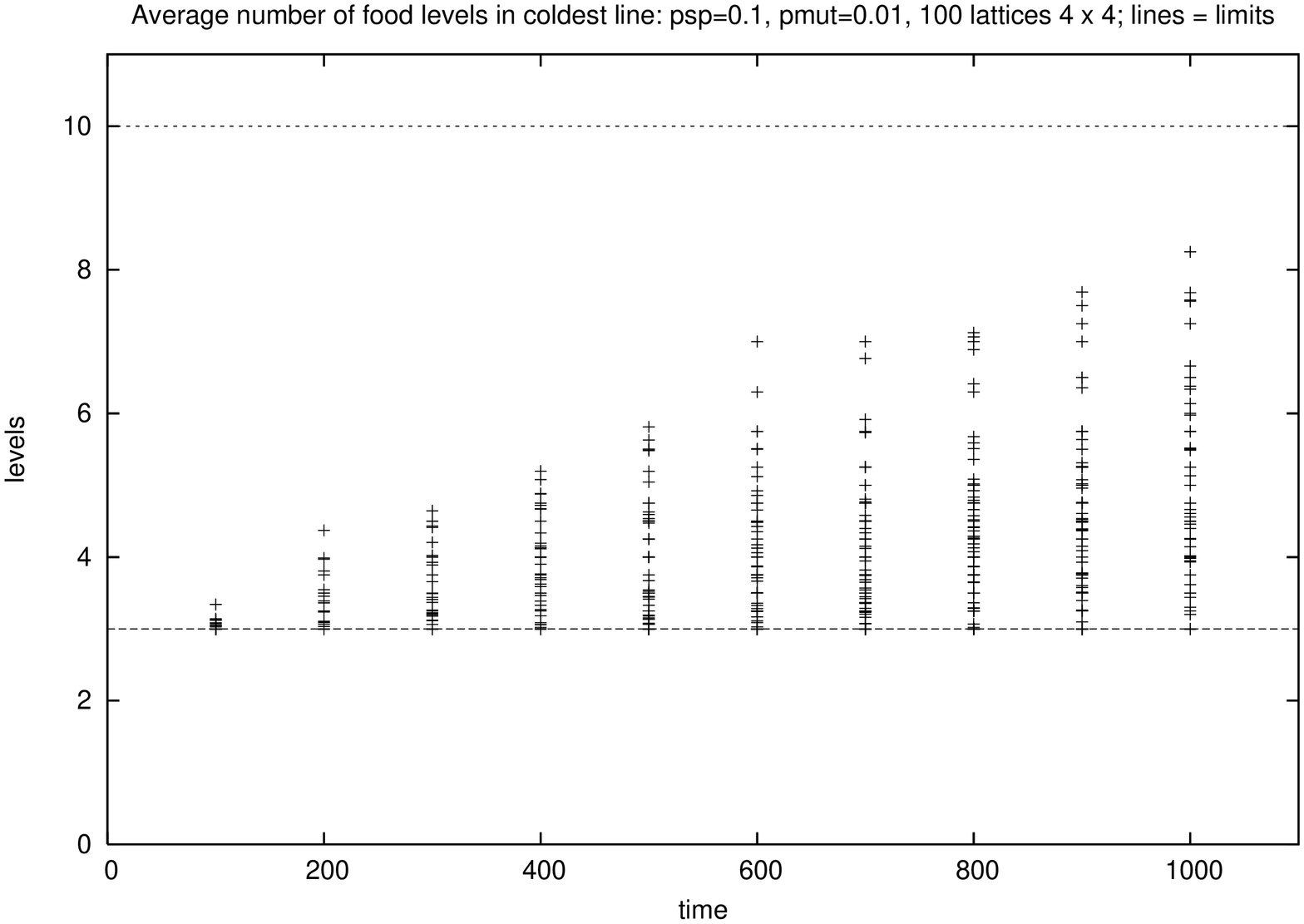}
\includegraphics[angle=-90,scale=0.31]{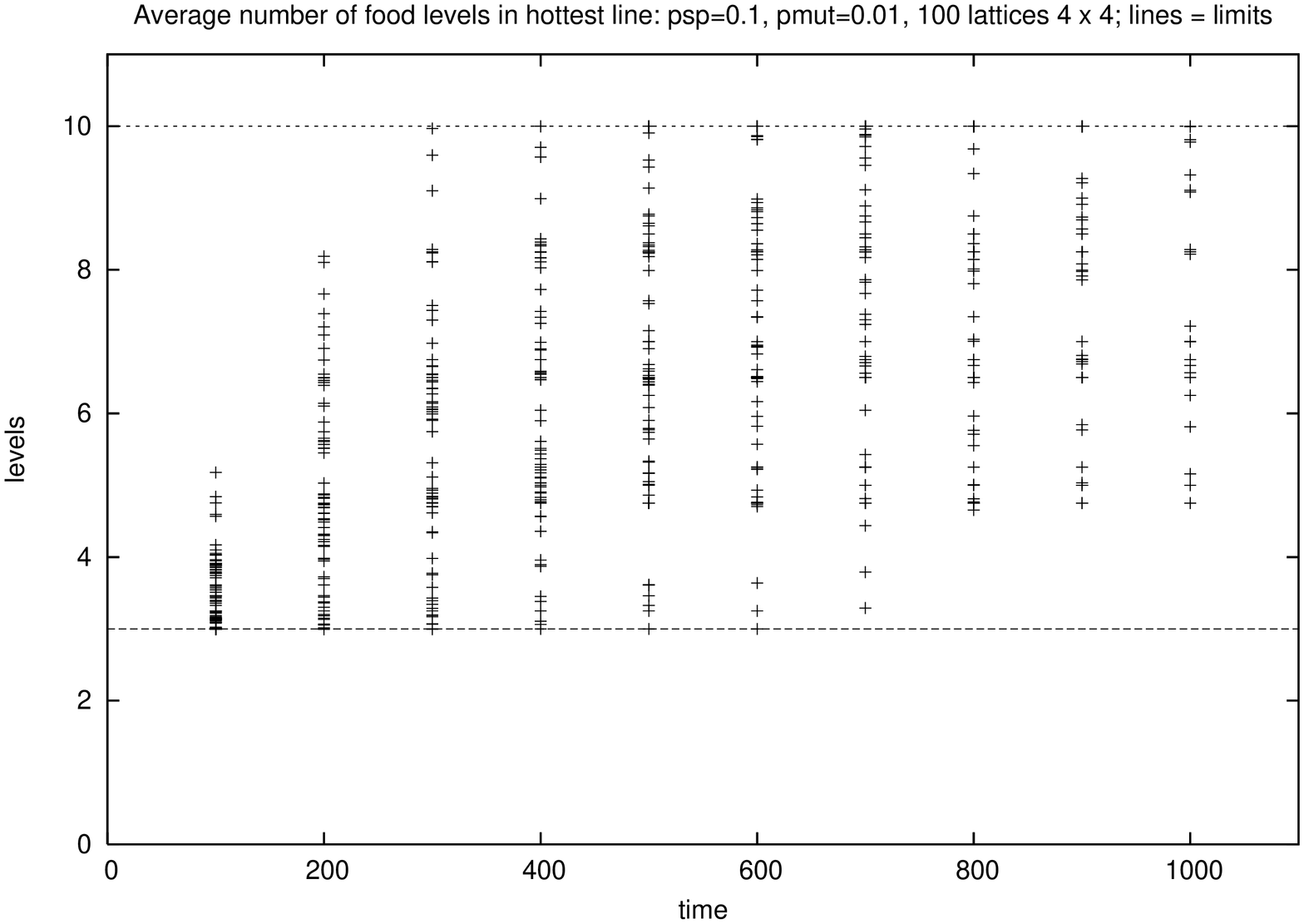}
\end{center}
\caption{
Time development in the number of food levels for the coldest (part a) and 
the hottest (part b) region. For 10 time intervals of 100 iterations each, we
show each of hundred samples, but the symbols may overlap. 
}
\end{figure}

\begin{figure}[hbt]
\begin{center}
\includegraphics[angle=-90,scale=0.5]{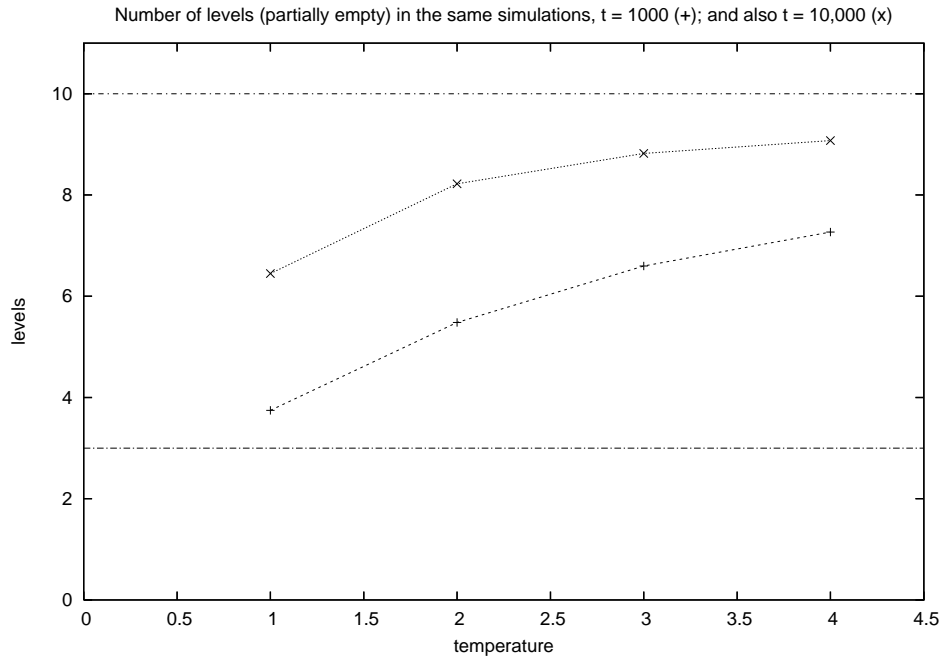}
\end{center}
\caption{From the simulations in Fig.6 and additional ones up to $t = 10^4$,
we show the temperature dependence of the average number of levels.
}
\end{figure}

\begin{figure}[hbt]
\begin{center}
\includegraphics[angle=-90,scale=0.5]{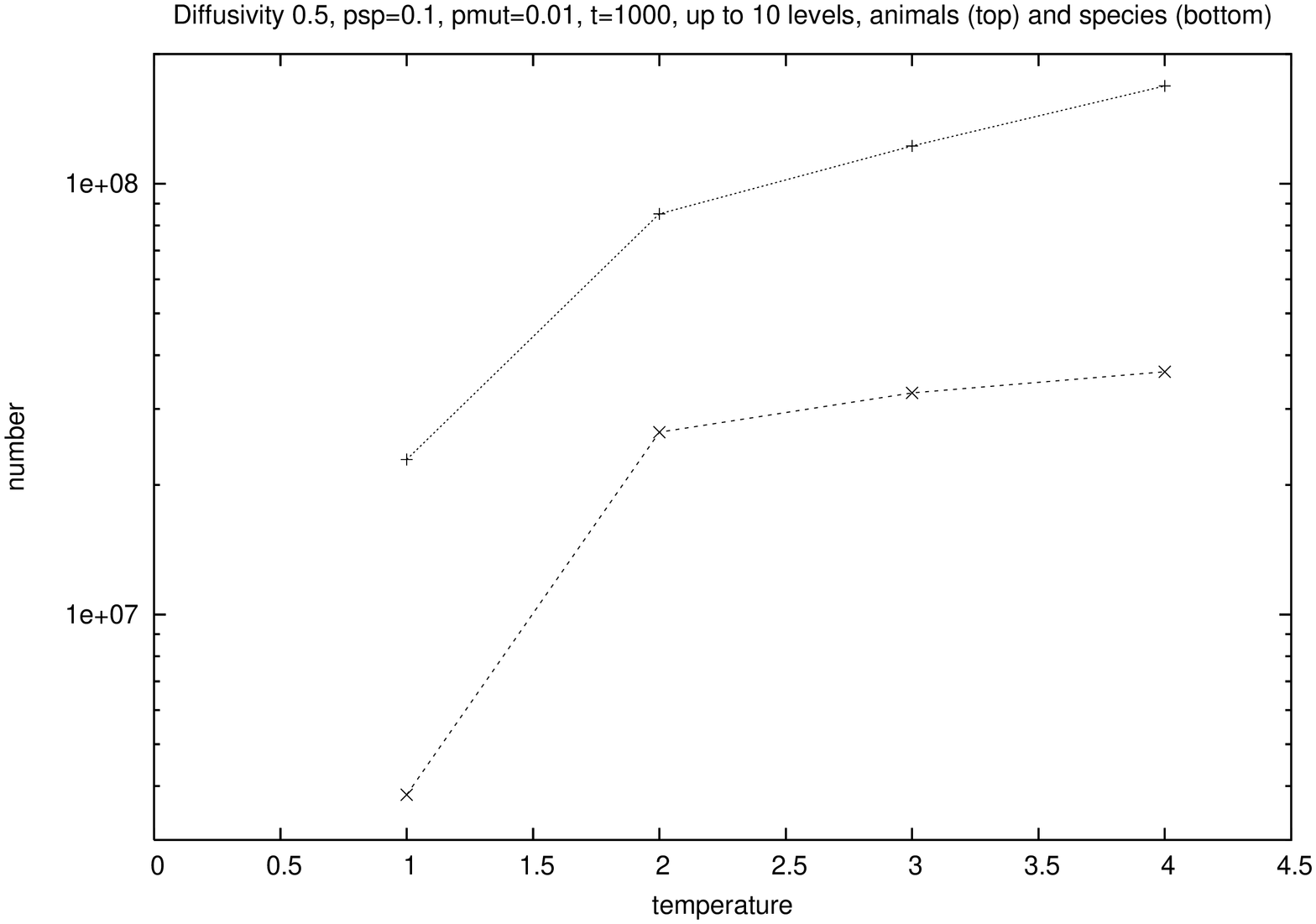}
\end{center}
\caption{Simulations as in Fig.7 but with diffusivity 1/2 instead of 
zero. 
}
\end{figure}

\section{Model and Results}

The Chowdhury model \cite{csk,Chowdhury,Stauffer} simulates the birth, ageing
and death of individual animals. The minimum reproduction age, number of
offspring per birth, the prey-predator relations and the number of food levels
self-organize, with a new food level added if the total number of animals is
not too large. Also mutation, speciation, and migration on a square lattice
are included. We refer to recent reviews \cite{Chowdhury,Stauffer} for details 
and a discussion of similar but simpler models, and to \cite{Kunwar} for a
systematic study of the dependence on various parameters. The present
simulations used a $4 \times 4$ lattice, up to $10^8$ time steps for each 
animal, and up to 10 food levels; food level $\ell$ allowed for up to
$2^{\ell-1}$ different species. If not given below otherwise, the mutation rate 
was $p_{mut} = 0.01$, while the carrying capacity $n_{max}$
(maximum number of animals per species and per lattice site) was 100.

Now we interpret the four lines of the lattice as cold, temperate, subtropic
and tropic latitudes, and assume the probability to give birth as proportional
to the line number. Thus in the tropics this probability is four times higher 
than in the cold regions. In the plots below we denote the line number as
temperature. Births in the Chowdhury model are offspring surviving until their
mortality becomes independent of age (below the minimum reproduction age). 

When this model was simulated for speciation rate $p_{sp} = 0.1$, the number of 
animals 
increased and the number of species decreased for increasing temperature. 
This does not correspond to what is usually found in nature. Then we omitted 
migration and kept the number of food levels fixed at six (or at ten), 
thus finding a slight increase of the number of 
species, much weaker than the increase for the number of animals. Finally we 
reduced $p_{sp}$ from 0.1 to 0.0001, and found the desired strong increase
in the numbers of both animals and species, Fig.1. The variation with 
speciation probability is shown in Fig.2; we see that it does not affect 
the temperature dependence of the animal numbers, but strongly
affects that of the species numbers. Only for low speciation probabilities,
when ecological niches left empty by an extinct species stay empty for a long
time, do we see a strong temperature dependence also for the species numbers.
The dependence on $p_{mut}$ is shown in Fig.3, that on $n_{max}= 100 \dots 
10000$ is not shown since the species numbers barely change.

In the above results, no migration was allowed. Including diffusion to
neighbouring empty lattice sites makes it easier to refill ecological niches
after extinctions and reduces the temperature dependence of the species 
numbers, similar to the variation with $p_{sp}$ in Fig.2. Fig.4 shows this 
effect for the case that the diffusion is proportional to the line number, i.e. 
our times higher in the tropical than in cold climates, and in the tropics
increases from 0 to 0.5 from bottom to top in Fig.4. Fig.5 shows the
species numbers if the diffusivity, instead of depending on the geographical
latitude as in Fig.4, is proportional to the height of the food level. Thus
the big predators on the top of the food web move more than the prey on the 
bottom. Instead of summing up over all levels, this Fig.5 shows the numbers 
in the various levels. The number of top predators now depends slightly less on 
temperature than that of the bottom prey. 

Until now we kept the numbers of food levels fixed, as in some earlier versions 
of the Chowdhury model; otherwise, as mentioned above we found a decrease
of species number with increasing temperature. If we allow a new food level
to be added, with probability 0.1, if the total number of animals on the 
lattice site is greater (and not smaller as before) than $n_{max}$, then 
we get species numbers increasing with temperature, similar to Fig.1 (not 
shown). Fig.6 shows the dependence of the number of levels on time, and Fig. 7 
on temperature. The number of levels soon reaches the maximum of ten which the
program does not allow to exceed. Migration was not allowed. 
(We start with 3 levels and do not allow more than 10 levels, as symbolized by 
the horizontal lines in Figs. 6 and 7.)

For this case of  increasing numbers of food levels and relatively
short time spans, a constant strong probability 1/2 of migrating to a
neighbouring empty niche  gave a rapid increase of species numbers
with temperature from cold to temperate (1 to 2) latitudes, followed
by a slight increase to warmer latitudes, as illustrated in Fig. 8.

\section{Discussion}

The two major aspects examined in this paper are latitudinal 
gradients in {\it species} numbers and in productivity, the latter measured 
as {\it animal} numbers.

Latitudinal gradients in species diversity, i.e., an (often very 
great) increase in species numbers from high to low latitudes, are 
the most distinct and best studied geographical trends.  There are 
exceptions, but most groups of animals and plants are much richer in 
the tropics than in temperate and cold climates, as reviewed in
\cite{Rohde92,Rohde99,Willig}. For example, the shallow waters of the 
Philippines have about 2000 fish species, South Australia has about 
300, and the Antarctic/Subantarctic Ocean less than 100 (further 
details and references in \cite{Rohde78}). Fossil evidence suggests that 
latitudinal diversity gradients have existed for at least 270 million 
years \cite{Stehli} or may even be time invariant \cite{Crame}.

The relationship between productivity and latitudinal gradients is by 
no means clear. For example, the Antarctic/Subantarctic Ocean is 
highly productive , as indicated for example by its huge quantities 
of krill, the staple food of many whales. On the other hand, most 
tropical seas do not have particularly high productivity but are very 
rich in species, although coral reefs  are both species rich and 
productive. The mean production in the open ocean in gram of carbon per
square meter, is 50 per year, 
that in the coastal zone 100 per year, whereas  a coral reef had a 
production of about 3500  per year (for details and references see 
\cite{Rohde98}). According to \cite{Waide}, who evaluated data 
from many studies, the relationship between diversity and 
productivity depends on scale. In this study we are concerned with 
continental/global, and with regional scales. At the former (more 
than 4000 km), unimodal or positive relationships are frequently 
found, although absence of or negative correlations occur as well. At 
the latter (200-4000 km) absence of or negative correlations are 
approximately as common as unimodal or positive relationships for 
plants, and slightly more common for animals.

Among the best known systems are freshwater ones. Approximately 40\% 
of fish species and almost 20\% of all vertebrates occur in freshwater 
(p. 126 in \cite{Myers}).
With regard to freshwater lakes, most species of fish are found in 
some large African tropical lakes. Lakes Victoria, Tanganyika and 
Malawi (121, 500 km$^2$) have a total of about 1,450 freshwater fish 
species (17\% of the Earth's total) (\cite{Myers}, p. 127). Lake Malawi 
(28, 231 km$^2$) alone contains 550 or more species. In contrast, the 
cold-temperate large lakes in the northern hemisphere are much 
poorer. Thus, only 173 fish species are found in the North American 
Great Lakes (246, 900 km$^2$) (\cite{Myers}, p. 127), and Lake Baikal has 
39 (Sheremetyev, personal communication), and this in spite of the 
North American Great Lakes and Lake Baikal together containing 31\% of 
the Earth's freshwater. The poor fauna cannot be due to glaciations. 
The high endemicity of its fauna indicates that Lake Baikal was 
little affected by them during the Ice Ages.

Productivity is generally higher in low than in high latitude lakes, 
with much overlap. Net primary productivity, measured in gram of carbon per
square meter,  of tropical lakes ranges 
from 0.100 to 7.6 per day, and from 30 to 2500  per year. In temperate lakes it
ranges from 0.005 to 3.6 per day, and from 2 to 950 per year 
(\cite{Likens}, p. 192). Annual productivity in the North American Great Lakes 
is ca. 100 to 310 (\cite{Wetzel}, Table 14-10), or 80-90 to 240-250 
(\cite{Likens}, pp. 194-195); in Lake Baikal it is 122.5 (\cite{Likens}, p. 
194). In contrast, Lake Victoria, has an annual production of 640 
(\cite{Wetzel}, Table 14-10), among the highest for freshwater lakes. Evidence 
is strong that the highest diversity of freshwater fishes is found in 
tropical, highly productive lakes.

The correlation between species diversity and productivity may often 
be complicated by geographical and/or historical contingencies (such 
as upwellings in oceans that carry inorganic substances to the 
surface resulting in increased productivity). Such contingencies 
cannot be considered in general evolutionary/ecological models, such 
as the Chowdhury model. However, experimental evidence suggests that 
there should indeed be an increase in productivity (measured as 
animal numbers in the model) with diversity, even if all other 
conditions are equal. Experimentally increasing biodiversity by a 
factor of two or three increased productivity by the same factor 
\cite{Kareiva}. Experimental studies reviewed in \cite{Hector} also 
showed that greater species diversity increases productivity. These 
studies provide evidence that primary productivity can be positively 
correlated with plant species richness for two reasons, differences 
between species in richer systems may allow complementary use of 
resources, and there is a higher probability of containing highly 
productive species in richer systems.

In summary, we return to the 5 aims stated at the end of the Introduction:
1) An increase of species numbers with temperature is reproduced in Figs.
1 to 5. 2) Figs. 6 and 7 show that the complexity of the foodwebs increases
with time and at a higher rate in the tropics. 
3) Since we did not allow for migration in Figs. 1 to 3, the 
increase of species with increasing temperature means also a higher creation 
rate of species in warmer climates. 4) Also the animal numbers increased,
usually even stronger that the species number, with increasing temperature.
5) The biodiversity gradient for fast-moving predators is lower than that for 
slow-moving prey, but only slightly.  

Therefore the simulations, qualitatively at least but
to some degree (as far as is possible in view of the great
variability of natural gradients) also quantitatively, agree with our aims.


\begin{thebibliography}{99}

\bibitem{csk}
Chowdhury, D, Stauffer D and Kunwar A. 2003 Unification of Small and 
Large Time Scales for Biological Evolution: Deviations from Power 
Law. Physical Review Letters 90, 068101.

\bibitem{Chowdhury}
Chowdhury, D. and Stauffer, D. 2005.  Evolutionary ecology in-silico:
Does mathematical modelling help in understanding the 'generic' trends?
J. Biosci. (India) 30, 277-287.

\bibitem{Stauffer}
Stauffer, D. and Chowdhury D. 2005.  Evolutionary ecology in-silico: 
evolving foodwebs, migrating population and speciation. Physica A 352,
202-215.

\bibitem{Jablonski99}
Jablonski, D. 1999. The future of the fossil record. Science 284, 2114-2116.

\bibitem{Rohde93}
Rohde, K. 1993 Ecology of Marine Parasites. 2nd edition. CAB - 
International, Wallingford, Oxon, U.K..

\bibitem{Rohde92}
Rohde, K. 1992.  Latitudinal gradients in species diversity:  the 
search for the primary cause. Oikos 65, 514 - 527.

\bibitem{Rohde05}
Rohde., K. 2005. Nonequilibrium Ecology. Cambridge University Press, 
Cambridge.

\bibitem{Jablonski93}
Jablonski, D. 1993. The tropics as a source of evolutionary novelty 
through geological time. Nature 364, 142-144.

\bibitem{Kunwar}
Kunwar, A. 2004. Evolution of spatially inhomogeneous eco-systems. A 
unified model based approach. Int. J. Mod. Phys. C 15, 1449-1460.

\bibitem{Rohde99}
Rohde, K. 1999. Latitudinal gradients in species diversity and
Rapoport's rule revisited: a review of recent work, and what can
parasites teach us about the causes of the gradients? Ecography, 22,
593-613.

\bibitem{Willig}
Willig, M.R. 2001. Latitude, common trends within. In "Encyclopedia 
of Biodiversity" vol.3 (ed.Levin, S.), pp. 701-714. Academic Press, 
N.Y.

\bibitem{Rohde78}
Rohde K. 1978. Latitudinal gradients in species diversity and their 
causes. II. Marine parasitological evidence for a time hypothesis. 
Biologisches Zentralblatt 97, 405-418.

\bibitem{Stehli}
Stehli, F.G., Douglas, D.G. and Newell, N.D. (1969). Generation and 
maintenance of gradients in taxonomic diversity. Science 164, 947-949.

\bibitem{Crame}
Crame, J.A. 2001. Taxonomic diversity gradients through geological
time. Diversity and Distributions 7, 175-189.

\bibitem{Rohde98}
Rohde, K. 1998. Latitudinal gradients in species diversity. Area 
matters, but how much? Oikos, 82, 184-190.

\bibitem{Waide}
Waide, R.B., Willig, M.R., Steiner, C.F., Mittelbach, G.G., Gough, 
L., Dodson, S.I., Juday, G.P. and Parmenter, R. 1999. The 
relationship between productivity and species richness. Annual Review 
of Ecology and Systematics 30, 257-300. 

\bibitem{Myers}
Myers, N. 1997. The niche diversity of biodiversity issues. In: 
Reaka-Kudla, M.L., Wilson, D.E. and Wilson, E.U. eds. Biodiversity 
II. Understanding and perfecting our biological resources. Joseph 
Henry Press, Washington, pp. 125-138.

\bibitem{Likens}
Likens, G.E. 1975. Primary production of inland ecosystems. In: 
Lieth, H. and Whittaker, R.H. eds. Primary productivity in the 
biosphere. Springer Verlag, Berlin, pp. 185-202.

\bibitem{Wetzel}
Wetzel, R.G. 1975. Limnology. Saunders, Philadelphia, PA.

\bibitem{Kareiva}
Kareiva, P. 1994. Diversity begets productivity. Nature 368, 686-687.

\bibitem{Hector}
Hector, A. 1998. The effect of diversity on productivity detecting
the role of species complementarity. Oikos 82, 597-599.

\end{thebibliography}
\end{document}